\documentclass[a4paper]{jpconf}
\usepackage{graphicx}
\begin{document}
\title{Quarkonium measurements via the di-muon decay channel in $\rm p$+$\rm p$ and $\rm Au$+$\rm Au$ collisions with the STAR experiment}

\author{Takahito Todoroki (for the STAR collaboration)}

\address{Brookhaven National Laboratory, Upton, New York 11973, USA}

\ead{todoroki@bnl.gov}


\begin{abstract}
We present the first $J/\psi$ and $\Upsilon$ measurements in the di-muon decay channel at mid-rapidity at RHIC using the newly installed Muon Telescope Detector.
In p$+$p collisions at $\sqrt{s}=500$~GeV, inclusive $J/\psi$ cross section can be described by CGC+NRQCD and NLO NRQCD model calculations for $0<p_T<20$~GeV/$c$.
In Au$+$Au collisions at $\sqrt{s_{NN}}=200$~GeV, we observe
(i) clear $J/\psi$ suppression indicating dissociation;
(ii) $J/\psi$ $R_{AA}$ can be qualitatively described by transport models including dissociation and regeneration with a tension at high $p_T$;
and (iii) hint of less melting of $\Upsilon(2S+3S)$ relative to $\Upsilon(1S)$ at RHIC compared to that at LHC. 
\end{abstract}

\section{Introduction}

Quarkonia are an essential probe to study the properties of the Quark Gluon Plasma (QGP).
The suppression of  $J/\psi$ due to color-screening effects in the medium was initially proposed as a direct evidence of the QGP formation~\cite{Matsui:1986dk}.
However, the interpretation of the $J/\psi$ suppression is still a challenge due to the contributions from the regenerated $J/\psi$ by
the recombination of $c\bar{c}$ pairs in the medium and the cold nuclear matter effects. Therefore it is important to have more precise
$J/\psi$ measurements over a broad kinematic range and even cleaner $\Upsilon$ state measurements. The latter do not suffer from the regeneration contribution due to the much
smaller $b\bar{b}$ pair cross section, i.e. $\sigma_{b\bar{b}}\sim2$~$\mu$b~\cite{Adare:2009ic} while $\sigma_{c\bar{c}}\sim800$~$\mu$b~\cite{Adamczyk:2012af} at top RHIC energy.
The newly installed Muon Telescope Detector (MTD), which provides both the di-muon trigger and the muon identification capability at mid-rapidity,
opens the door to measuring quarkonia via the di-muon decay channel at STAR.
Compared to the di-electron decay channel, the di-muon decay channel suffers much less from bremsstrahlung and thus provides
much better invariant mass resolution to separate different $\Upsilon$ states.
Using the MTD di-muon trigger, the STAR experiment recorded data corresponding to an integrated luminosity of 28.3~$pb^{-1}$ in p$+$p collisions at $\sqrt{s}=500$~GeV in the RHIC 2013 run,
and 14.2~$nb^{-1}$ in Au$+$Au collisions at $\sqrt{s_{NN}}=200$~GeV in the RHIC 2014 run.
In these proceedings, we report (1) the measurements of $J/\psi$ production in p+p collisions at $\sqrt{s}=500$~GeV;
and  (2) the measurements of the nuclear modification factor ($R_{AA}$) for $J/\psi$ and the production of different $\Upsilon$ states in Au+Au collisions at $\sqrt{s_{NN}}=200$~GeV.

\section{$J/\psi$ measurements in p$+$p collisions at $\sqrt{s}=500$~GeV }
\begin{figure}[h]
\begin{minipage}{18pc}
\center
\includegraphics[width=18pc]{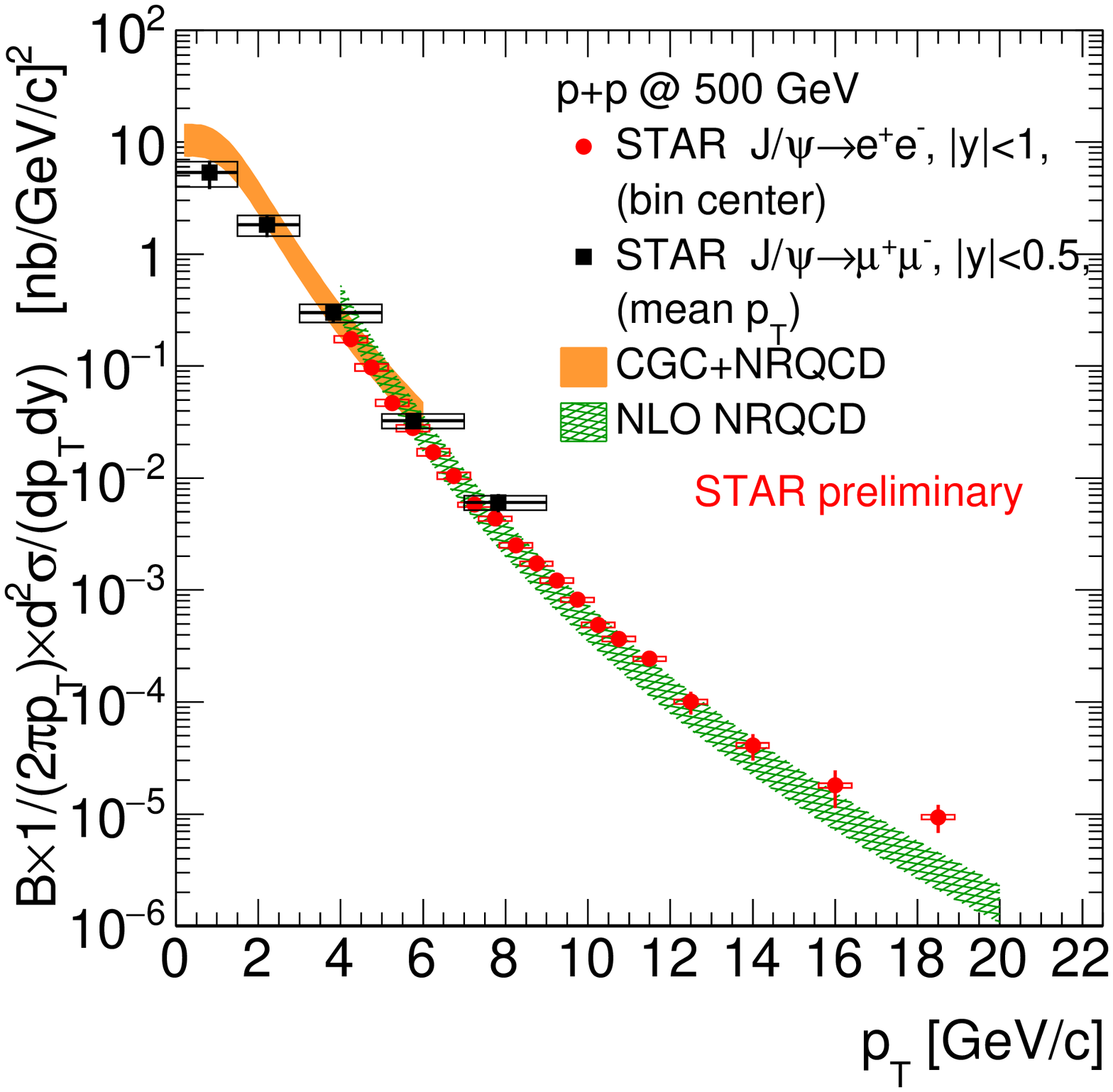}
\caption{\label{fig:run13JpsiCrossSection} $J/\psi$ cross section scaled by the branching ratio $B$ as a function of $p_T$ in the di-muon decay channel (black circle) 
and  in the di-electron decay channel (red circle).}
\end{minipage}\hspace{2pc}%
\begin{minipage}{18pc}
\center
\includegraphics[width=18pc]{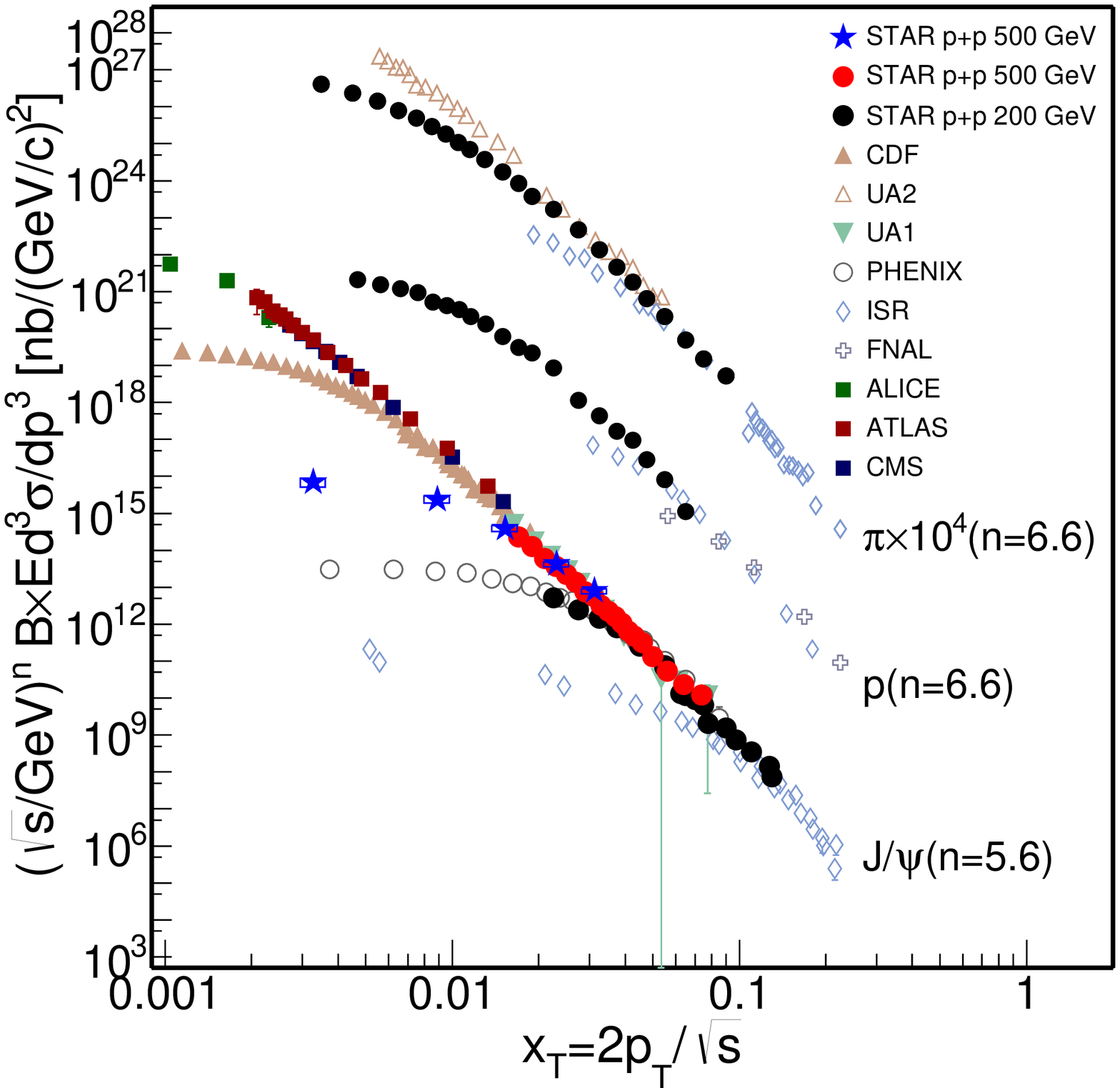}
\caption{\label{fig:run13XtScaling} $x_T$ scaling of $J/\psi$ cross section scaled by the branching ratio $B$ in the di-muon decay channel (blue star) and in the di-electron decay channel (red circle).}
\end{minipage} 
\end{figure}

Figure~\ref{fig:run13JpsiCrossSection} shows the cross section of $J/\psi$ in p$+$p collisions at $\sqrt{s}=500$~GeV in the di-electron
and di-muon decay channels for $0<p_T<20$~GeV/$c$. The di-muon decay channel extends $p_T$ reach down to 0~GeV/$c$.
The results in these decay channels are consistent in the overlapping $p_T$ range of $4<p_T<9$~GeV/$c$.
The experimental results can be well described by CGC+NRQCD calculations at low $p_T$~\cite{Ma:2014mri}
and NLO NRQCD calculations at high $p_T$~\cite{Shao:2014yta}. Figure~\ref{fig:run13XtScaling} shows 
the $x_T=2p_T/\sqrt{s}$ scaling of $J/\psi$ cross section~\cite{Abelev:2009qaa}.
The $J/\psi$ cross section in p$+$p collisions at $\sqrt{s}=500$~GeV follows the common  trend as a function of $x_T$ at high $p_T$.
The breaking of the $x_T$ scaling at low $p_T$ can be attributed to the soft processes.

\section{$J/\psi$ measurements in Au$+$Au collisions at $\sqrt{s_{NN}}=200$~GeV }
\begin{figure}[h]
\begin{minipage}{18pc}
\center
\includegraphics[width=18pc]{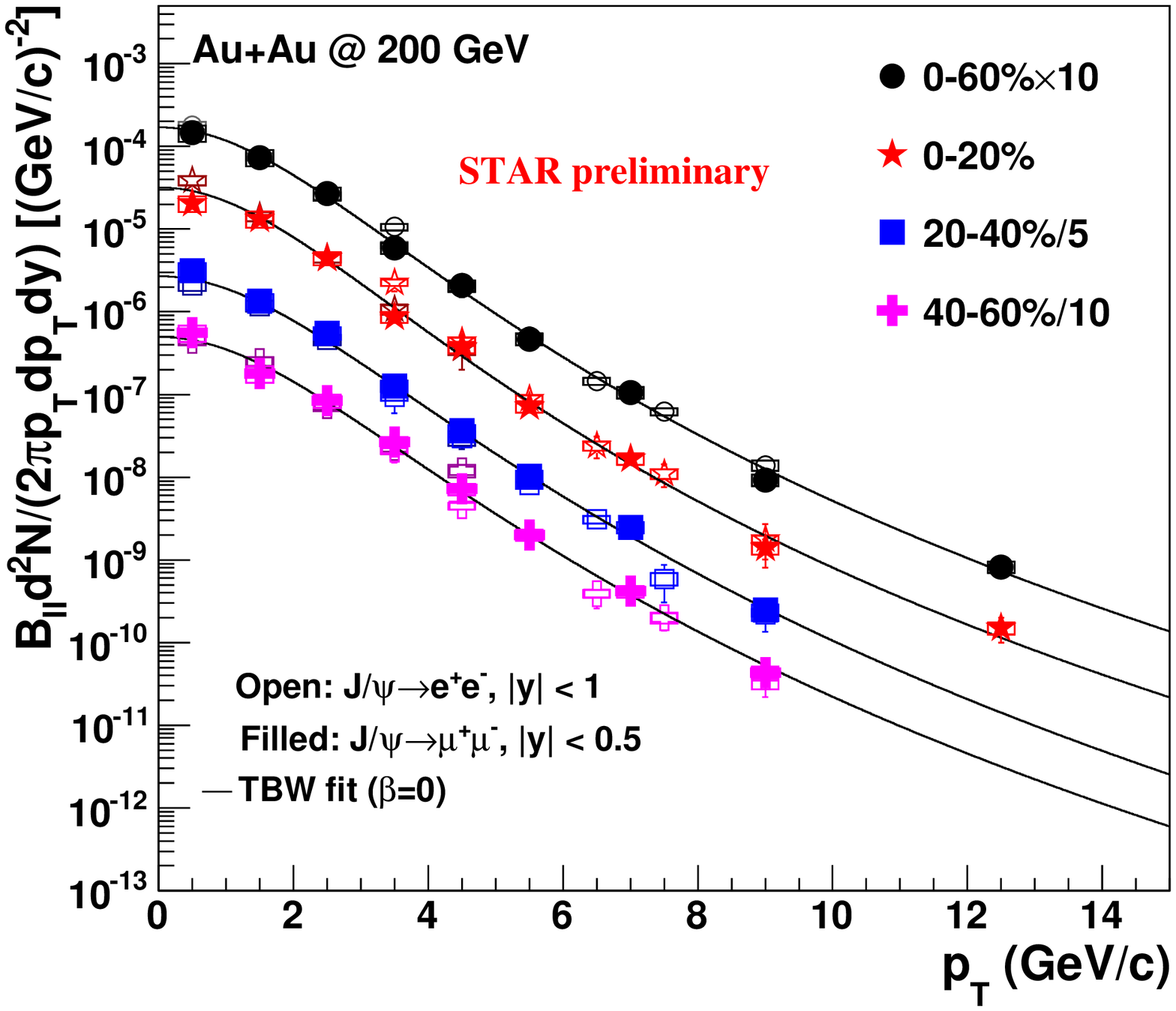}
\caption{\label{fig:JpsiYieldPt} Invariant yield of $J/\psi$ scaled by the branching ratio $B_{ll}$ as a function of $p_T$ in different centralities in the di-muon decay channel (filled) 
and in the di-electron decay channel (open) . 
}
\end{minipage} \hspace{2pc}%
\begin{minipage}{18pc}
\center
\includegraphics[width=18pc]{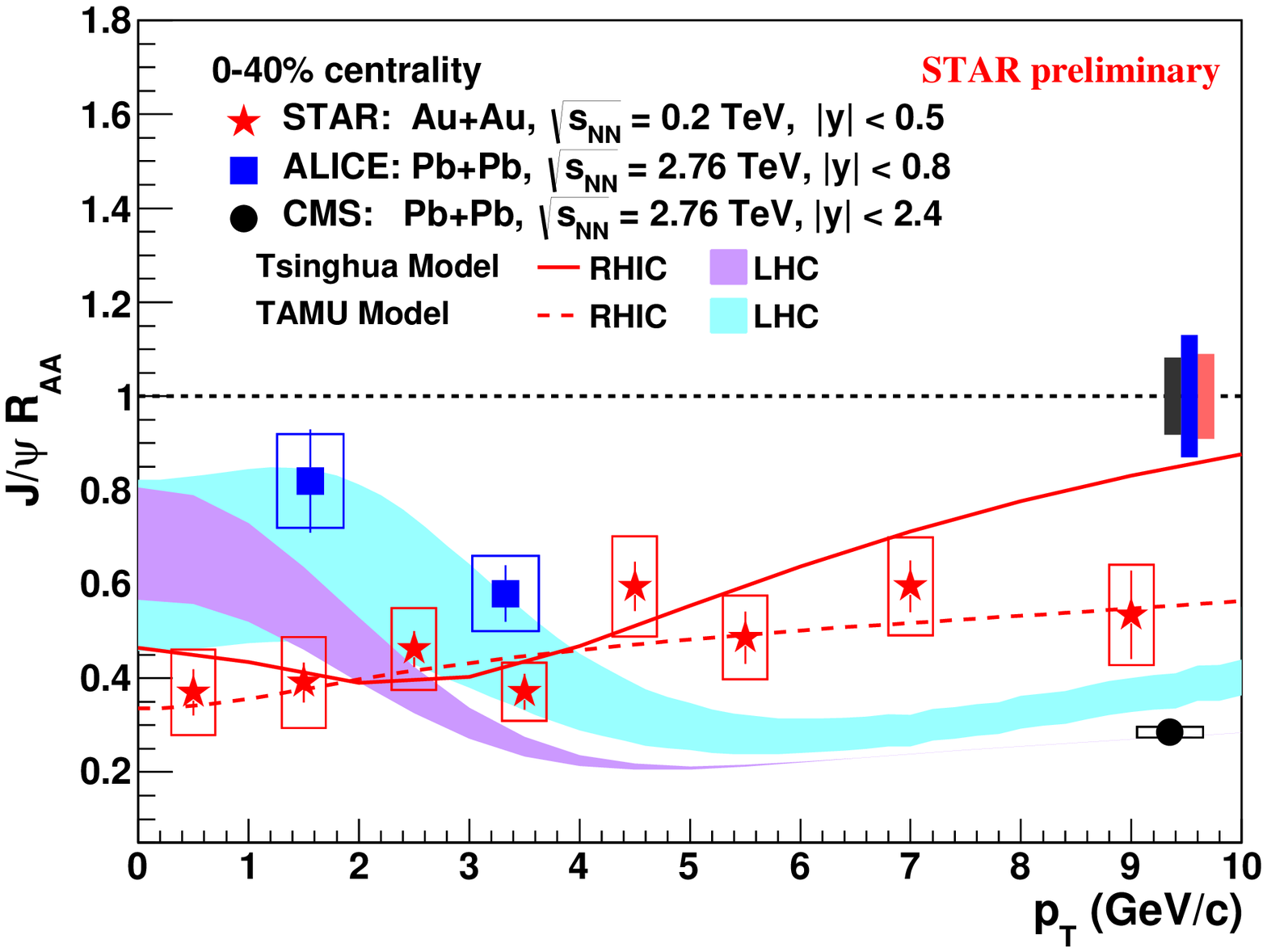}\hspace{2pc}%
\caption{\label{fig:JpsiRaaVsPt} $R_{AA}$ as a function of $p_T$ at RHIC (red star) 
and at LHC (blue square, black circle).
 The lines and bands indicate Transport Model calculations for RHIC and LHC energies.}
\end{minipage}
\end{figure}

Figure~\ref{fig:JpsiYieldPt} shows the invariant yield of $J/\psi$ in Au$+$Au collisions at $\sqrt{s_{NN}}=200$~GeV for different collision centralities.
The new results in the di-muon decay channel are consistent with previous results in the di-electron decay channel~\cite{Adamczyk:2012ey,Adamczyk:2013tvk} within uncertainties.

The nuclear modification factor $R_{AA}=\frac{\sigma_{inel}}{\left<N_{coll}\right>}\frac{d^2N_{AA}/dydp_T}{d^2\sigma_{pp}/dydp_T}$ of $J/\psi$
in 0-40\% central Au$+$Au collisions is shown in Fig.~\ref{fig:JpsiRaaVsPt}, compared with LHC results~\cite{Abelev:2013ila,Chatrchyan:2012np}.
The strong suppression at RHIC at low $p_T$ indicates that dissociation plays a significant role in this $p_T$ range. 
The hint of the increasing trend of $R_{AA}$ at RHIC at high $p_T$ can be explained by formation-time effects and feed-down of $B$ hadrons.
The less suppression of $J/\psi$ at LHC at low $p_T$ indicates larger regeneration contribution due to higher charm cross section,
while more suppression of $J/\psi$ at LHC at high $p_T$ indicates larger dissociation rate due to higher temperature of the medium. 
Transport Models from Tsinghua~\cite{Liu:2009nb, Zhou:2014kka} and Texas A\&M University (TAMU) ~\cite{Zhao:2010nk, Zhao:2011cv}, including dissociation and regeneration effects,
can qualitatively describe the $p_T$ dependence of RHIC and LHC data. 

\begin{figure}[h]
\begin{minipage}{18pc}
\center
\includegraphics[width=18pc]{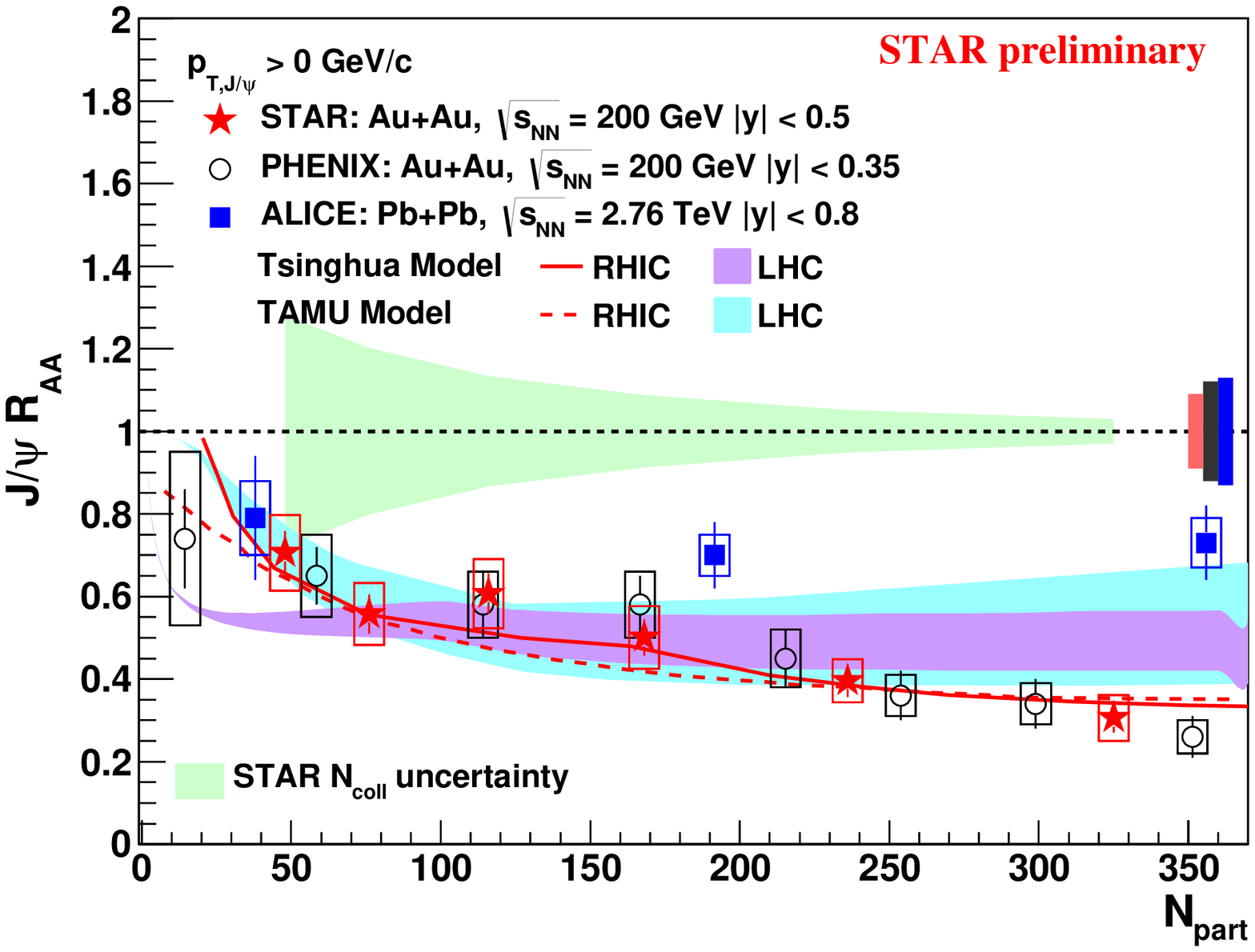}
\caption{\label{fig:JpsiRaaNpartLowPt} Nuclear modification factor $R_{AA}$ for integrated $p_T$ as a function of $N_{part}$.}
\end{minipage} \hspace{2pc}%
\begin{minipage}{18pc}
\center
\includegraphics[width=18pc]{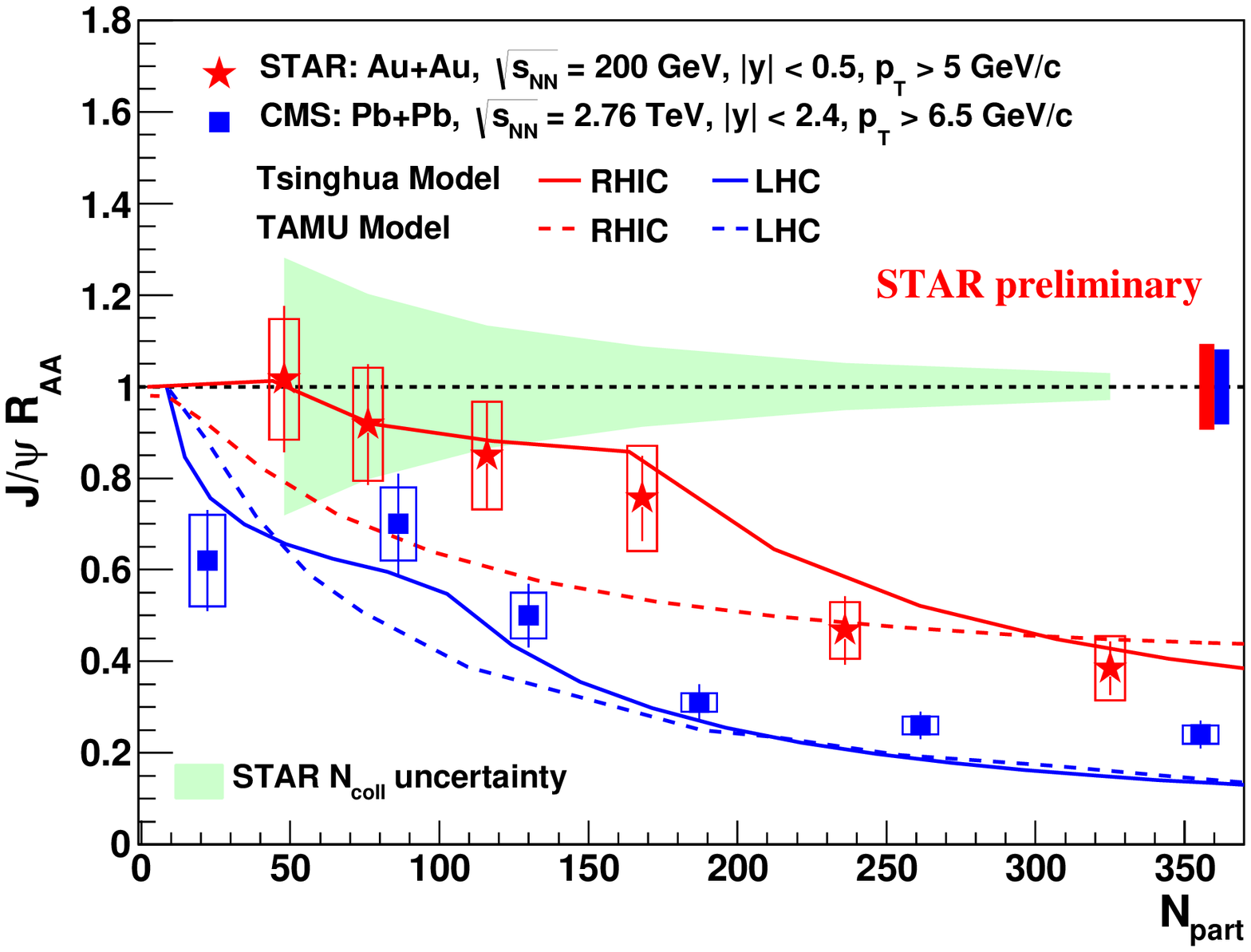}
\caption{\label{fig:JpsiRaaNpartHighPt} Nuclear modification factor $R_{AA}$ for $p_T>5$~GeV/$c$ as a function of $N_{part}$.}
\end{minipage}
\end{figure}

Centrality dependence of $J/\psi$ cross section is shown in Fig.~\ref{fig:JpsiRaaNpartLowPt} for integrated $p_T$ and in Fig.~\ref{fig:JpsiRaaNpartHighPt} for $p_T>5$~GeV/$c$.
For integrated $p_T$, both models can describe centrality dependence at RHIC, but tend to overestimate suppression at LHC.
For $p_T>5$~GeV/$c$, there is tension among models and data. New measurements in the di-muon decay channel  provide a distinguishing power for these transport models.

\section{$\Upsilon$ measurements in Au$+$Au collisions at $\sqrt{s_{NN}}=200$~GeV }
\begin{figure}[h]
\begin{minipage}{18pc}
\center
\includegraphics[width=18pc]{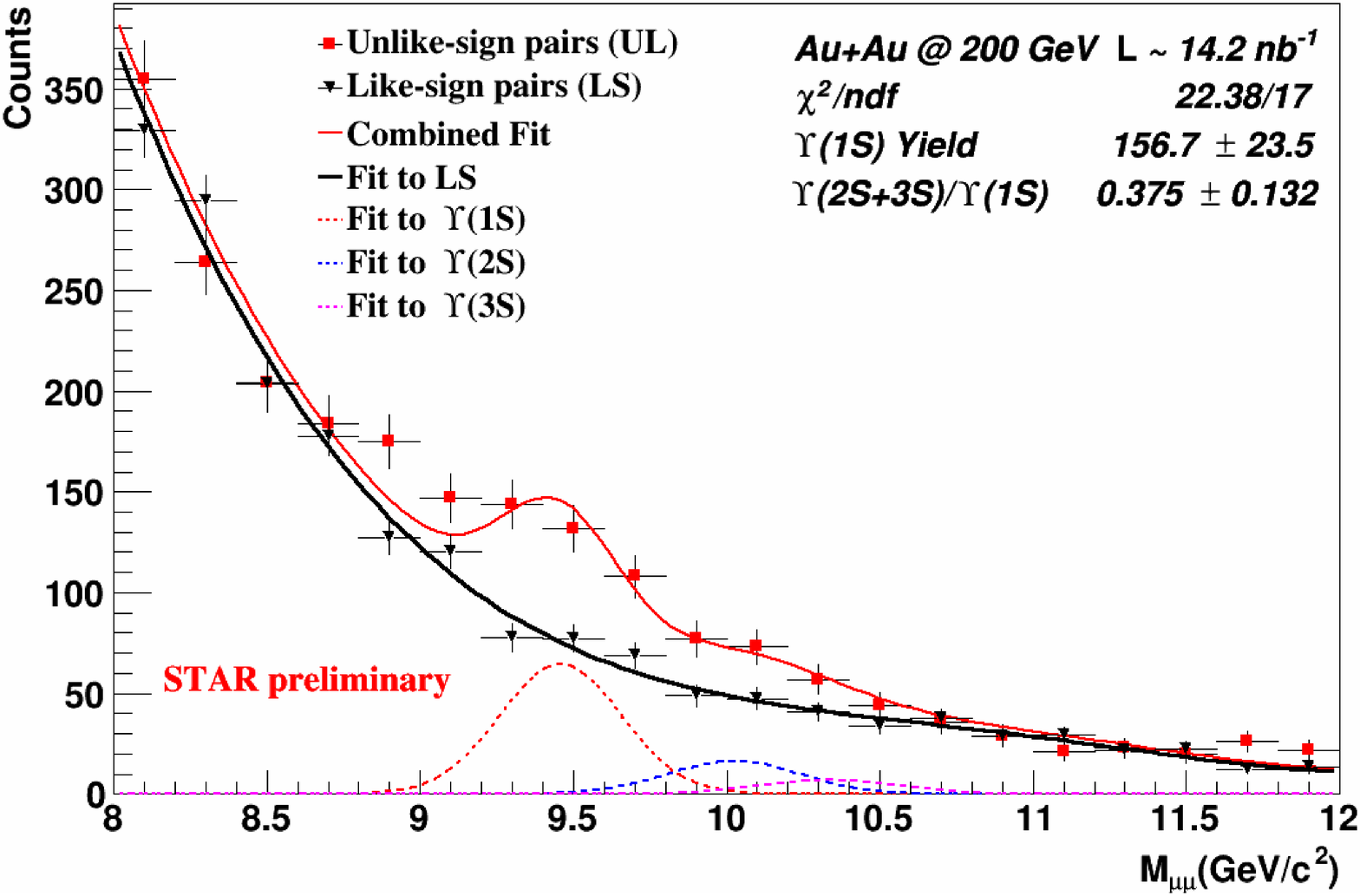}
\caption{\label{fig:Run14UpsSignal} Di-muon mass spectrum in the $\Upsilon$ state mass range.
}
\end{minipage}\hspace{2pc}%
\begin{minipage}{18pc}
\center
\includegraphics[width=18pc]{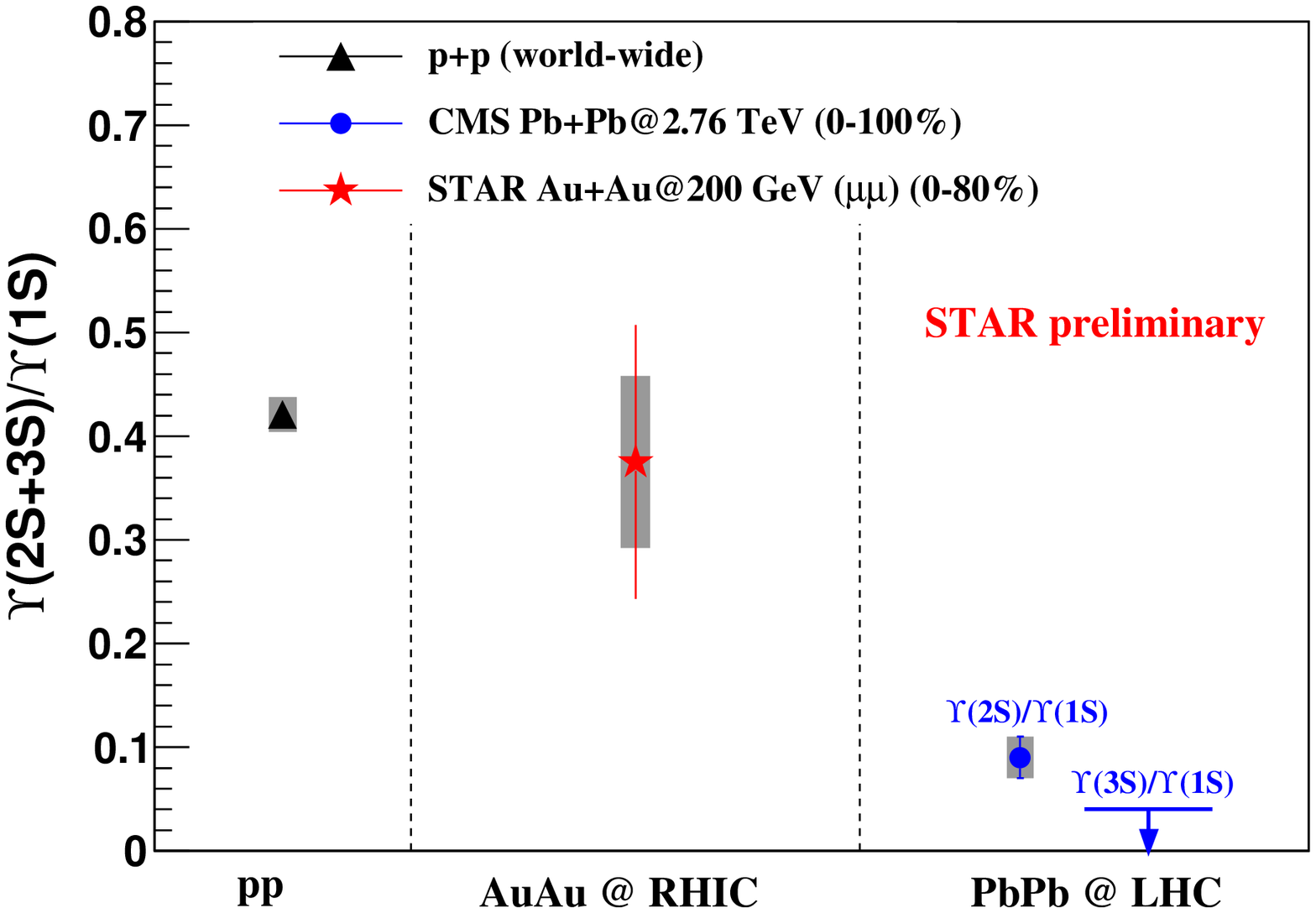}
\caption{\label{fig:UpsilonRatio} $\Upsilon(2S+3S)/\Upsilon(1S)$ ratio for world-wide p+p data, and for heavy-ion collisions at RHIC, and LHC energies.
}
\end{minipage} 
\end{figure}

Figure~\ref{fig:Run14UpsSignal} shows the di-muon mass spectrum in $\Upsilon$ state mass range in Au$+$Au collisions at $\sqrt{s_{NN}}=200$~GeV.
We observe signs of an indication of $\Upsilon(2S+3S)$ signals in the di-muon decay channel.
The raw yields of $\Upsilon$ states are obtained by a simultaneous fit to the like-sign and unlike-sign distributions.
In this fit, (i) the $\Upsilon$ state masses are fixed to the PDG values and their widths are determined by simulation;
(ii) the ratio of $\Upsilon(2S)/\Upsilon(3S)$ is fixed to the value in p$+$p collisions; and (iii) the shape of $b\bar{b}$ and Drell-Yan background is estimated using PYTHIA.
Figure~\ref{fig:UpsilonRatio} shows the fitted $\Upsilon(2S+3S)/\Upsilon(1S)$ ratio compared with the world-wide p+p data~\cite{Zha:2013uoa} and 
CMS data~\cite{Chatrchyan:2012lxa,Chatrchyan:2013nza}.
There is a hint of less melting of $\Upsilon(2S+3S)$ relative to $\Upsilon(1S)$ at RHIC than at LHC.

\section{Summary and Outlook}
We present the first $J/\psi$ and $\Upsilon$ measurements in the di-muon decay channel at mid-rapidity at RHIC.
In p$+$p collisions at $\sqrt{s}=500$~GeV, inclusive $J/\psi$ cross section can be described by CGC+NRQCD and NLO NRQCD model calculations for $0<p_T<20$~GeV/$c$.
In Au$+$Au collisions at $\sqrt{s_{NN}}=200$~GeV, we observe
(i) clear $J/\psi$ suppression indicating dissociation;
(ii) $J/\psi$ $R_{AA}$ can be qualitatively described by transport models including dissociation and regeneration despite a tension at high $p_T$;
and (iii) there is a hint of less melting of $\Upsilon(2S+3S)$ relative to $\Upsilon(1S)$ at RHIC compared to that at LHC. 
These measurements in Au$+$Au collisions will have better statistical precision by combining the similar amount of data recorded in the RHIC 2016 run.


\section*{References}
\begin{thebibliography}{20}
\bibitem{Matsui:1986dk} T. Matsui and H. Satz, {\it Phys. Lett. B} {\bf 178} 416-422 (1986)
\bibitem{Adare:2009ic} A. Adare {\it et. al.} (PHENIX Collaboration), {\it Phys. Rev. Lett.} {\bf 103} 082002 (2009)
\bibitem{Adamczyk:2012af} L. Adamczyk {\it et. al.} (STAR Collaboration), {\it Phys. Rev. D} {\bf 86} 072013 (2012)
\bibitem{Ma:2014mri} Yan-Qing Ma and Raju Venugopalan, {\it Phys. Rev. Lett.} {\bf 113} 192301 (2015)
\bibitem{Shao:2014yta} H. Shao {\it et. al.}, {\it JHEP} {\bf 05} 103 (2015)
\bibitem{Abelev:2009qaa} B. Abelev {\it et. al.} (STAR Collaboration), {\it Phys. Rev. C} {\bf 80} 041902 (2009)
\bibitem{Adamczyk:2012ey} L. Adamczyk {\it et. al.} (STAR Collaboration), {\it Phys. Let. B} {\bf 722} 55-62 (2013)
\bibitem{Adamczyk:2013tvk} L. Adamczyk {\it et. al.} (STAR Collaboration), {\it Phys. Rev. C} {\bf 90} 024906 (2014)
\bibitem{Liu:2009nb} Y. Liu {\it et. al.}, {\it Phys. Let. B} {\bf 678} 72-76 (2009)
\bibitem{Zhou:2014kka} K. Zhou {\it et. al.}, {\it Phys. Rev. C} {\bf 89} 054911 (2014)
\bibitem{Zhao:2010nk} X. Zhao and R. Rapp, {\it Phys. Rev. C} {\bf 82} 064905 (2010)
\bibitem{Zhao:2011cv} X. Zhao and R. Rapp, {\it Nucl. Phys. A} {\bf 859} 114-125 (2011)
\bibitem{Abelev:2013ila} B. Abelev {\it et. al.} (ALICE Collaboration), {\it Phys. Let. B} {\bf 734} 314-327 (2014)
\bibitem{Chatrchyan:2012np} S. Chatrchyan {\it et. al.} (CMS Collaboration), {\it JHEP} {\bf 05} 063 (2012)
\bibitem{Adamczyk:2012pw} L. Adamczyk {\it et. al.} (STAR Collaboration), {\it Phys. Rev. Lett.} {\bf 111} 052301 (2013)
\bibitem{Zha:2013uoa} W. Zha {\it et. al.}, {\it Phys. Rev. C} {\bf 88} 067901 (2013)
\bibitem{Chatrchyan:2012lxa} S. Chatrchyan {\it et. al.} (CMS Collaboration), {\it Phys. Rev. Lett.} {\bf 109} 222301 (2012)
\bibitem{Chatrchyan:2013nza} S. Chatrchyan {\it et. al.} (CMS Collaboration), {\it JHEP} {\bf 04} 103 (2014)
\end{thebibliography}
\end{document}